\documentclass[aps,prl,amsmath,amssymb,showpacs,superscriptaddress,nofootinbib,twocolumn]{revtex4-1}
\usepackage{multirow}
\usepackage{float}
\usepackage{graphicx}% Include figure files
\usepackage{dcolumn}% Align table columns on decimal point
\usepackage{bm}% bold math
\usepackage{rotating}
\usepackage{color}
\usepackage{subfigure}
\usepackage{pstricks}
\usepackage{soul}
\usepackage{overpic}
\usepackage[english]{babel}
\usepackage[colorlinks,linkcolor=blue,anchorcolor=blue,citecolor=blue]{hyperref}
\usepackage{tabularx}
\usepackage{comment}
\newcommand{\BR}{{\cal B}}
\newcommand{\azz}{a^{0}_{0}(980)}
\newcommand{\fz}{f_{0}(980)}

\newcommand{\psip}{\psi(3686)}
\newcommand{\jpsi}{J/\psi}

\newcommand{\chico}{\chi_{c1}}

\newcommand{\pip}{\pi^+}
\newcommand{\pim}{\pi^-}
\newcommand{\piz}{\pi^0}
\newcommand{\pppm}{\pi^{+}\pi^{-}}

\newcommand{\kpkm}{K^{+}K^{-}}
\newcommand{\kkbar}{K\bar{K}}

\newcommand{\gam}{\gamma}

\newcommand{\etap}{\eta^{\prime}}
\newcommand{\B}{\mathcal{B}}
\newcommand{\mulR}{\multirow}
\newcommand{\mulC}{\multicolumn}

\newcommand{\atof}{\azz$$\to$$\fz}
\newcommand{\ftoa}{\fz$$\to$$\azz}
\def \mevcc      {\,\mathrm{MeV}/c^2}

\def \gevcc      {\,\mathrm{GeV}/c^2}

\begin{document}

%\normalsize
%\parskip=0pt plus 1pt minus 1pt
\title{\boldmath Observation of $a^{0}_{0}(980)$-$f_{0}(980)$ Mixing}

\author{
\begin{small}
\begin{center}
M.~Ablikim$^{1}$, M.~N.~Achasov$^{9,d}$, S. ~Ahmed$^{14}$, M.~Albrecht$^{4}$, M.~Alekseev$^{53A,53C}$, A.~Amoroso$^{53A,53C}$, F.~F.~An$^{1}$, Q.~An$^{50,40}$, J.~Z.~Bai$^{1}$, Y.~Bai$^{39}$
% ...
\\
\vspace{0.2cm}
(BESIII Collaboration)\\
\vspace{0.2cm} {\it
$^{1}$ Institute of High Energy Physics, Beijing 100049, People's Republic of China\\
% ...
$^{4}$ Bochum Ruhr-University, D-44780 Bochum, Germany\\
% ...
$^{9}$ G.I. Budker Institute of Nuclear Physics SB RAS (BINP), Novosibirsk 630090, Russia\\
% ...
$^{14}$ Helmholtz Institute Mainz, Johann-Joachim-Becher-Weg 45, D-55099 Mainz, Germany\\
% ...
$^{39}$ Southeast University, Nanjing 211100, People's Republic of China\\
$^{40}$ State Key Laboratory of Particle Detection and Electronics, Beijing 100049, Hefei 230026, People's Republic of China\\
% ...
$^{50}$ University of Science and Technology of China, Hefei 230026, People's Republic of China\\
% ...
$^{53}$ (A)University of Turin, I-10125, Turin, Italy; (B)University of Eastern Piedmont, I-15121, Alessandria, Italy; (C)INFN, I-10125, Turin, Italy\\
% ...
$^{d}$ Also at the Novosibirsk State University, Novosibirsk, 630090, Russia\\
% ...
}
\end{center}
\end{small}
}
\noaffiliation{}

\date{\today}

\begin{abstract}

 We report the first observation of $a^{0}_{0}(980)$-$f_{0}(980)$ mixing in the decays of $J/\psi\to\phi f_{0}(980)\to\phi a^{0}_{0}(980)\to\phi\eta\pi^{0}$ and $\chi_{c1}\to a^{0}_{0}(980)\pi^{0}\to f_{0}(980)\pi^{0}\to\pi^{+}\pi^{-}\pi^{0}$, using data samples of $1.31\times10^{9}$ $J/\psi$ events and $4.48\times10^{8}$ $\psi(3686)$ events accumulated with the BESIII detector. The signals of $f_{0}(980)\to a^{0}_{0}(980)$ and $a^{0}_{0}(980)\to f_{0}(980)$ mixing are observed at levels of statistical significance of $7.4\sigma$ and $5.5\sigma$, respectively. The corresponding branching fractions and mixing intensities are measured and the constraint regions on the coupling constants, $g_{a_{0}K^{+}K^{-}}$ and $g_{f_{0}K^{+}K^{-}}$, are estimated. The results improve the understanding of the nature of $a^{0}_{0}(980)$ and $f_{0}(980)$.

\end{abstract}

\maketitle

 Since the discoveries of $\azz$ and $\fz$ several decades ago, explanations about the nature of these two light scalar mesons have been controversial. These two states, with similar masses but different decay modes and isospins, are difficult to accommodate in the traditional quark-antiquark model~\cite{qq}, and many alternative formulations have been proposed to explain their internal structure, including tetra-quarks~\cite{qq,qqqq}, $K\bar{K}$ molecule~\cite{KK}, or quark-antiquark gluon hybrid~\cite{gluon}.

 The mixing mechanism in the system of $\azz$-$\fz$, which was first proposed in the late 1970s~\cite{Achasov1979}, is thought to be an essential approach to clarify the nature of these two mesons. Both $\azz$ and $\fz$ can decay into $\kpkm$ and $K^0\bar{K^0}$, which show a difference of 8~MeV/$c^2$ in the production mass threshold due to isospin breaking effects. The mixing amplitude between $\azz$ and $\fz$ is dominated by the unitary cuts of the intermediate two-kaon system and proportional to the phase-space difference between them. As a consequence, a narrow peak of about 8~MeV/$c^2$ in width is predicted between the charged and neutral $K\bar{K}$ mass thresholds, while the normal widths of $\azz$ and $\fz$ should be $50-100$~MeV$/c^2$~\cite{PDG}. The mixing mechanism has been studied extensively in various aspects, and many reactions have been discussed, such as $\gamma p\to p\piz\eta$~\cite{Kerbikov}, $\pim p \to \piz\eta n$~\cite{Achasov20041,Achasov20042}, $p n \to d \piz\eta$~\cite{Kudr1,Kudr2,Kudr3}, $d d \to \alpha\piz\eta$~\cite{Grishina}. However, no quantitative experimental result has been firmly established yet.

 Inspired by Refs.~\cite{Close1, Close2, Hanhart}, a first quantitative calculation was carried out to examine the $\azz\leftrightarrow\fz$ mixing with the isospin-violating processes of $\jpsi\to\phi\fz\to\phi\azz\to\phi\eta\piz$ and $\chi_{c1}\to\piz\azz\to\piz\fz\to\piz\pip\pim$~\cite{JJWu1, JJWu2, Roca, Bayar}. The central masses and couplings of $\azz\to\eta\piz/\kkbar$ and $\fz\to\pi\pi/\kkbar$ from various models~\cite{qq,qqqq,KK,gluon} and different experimental results~\cite{CB,BNL,KLOE1,KLOE2,SND1,SND2} were investigated. The mixing intensities, \emph{i.e.}, $\xi_{fa}$ for the $\fz\to\azz$ transition and $\xi_{af}$ for the $\azz\to\fz$ transition, are defined as
 \begin{small}
 \begin{eqnarray}
   \xi_{fa}&=&\frac{\BR(\jpsi\to\phi\fz\to\phi\azz\to\phi\eta\piz)}{\BR(\jpsi\to\phi\fz\to\phi\pi\pi)}, \\
   \xi_{af}&=&\frac{\BR(\chi_{c1}\to\piz\azz\to\piz\fz\to\piz\pip\pim)}{\BR(\chi_{c1}\to\piz\azz\to\piz\piz\eta)}.
   \label{intensity}
 \end{eqnarray}
 \end{small}

 The mixing intensities, $\xi_{fa}$ and $\xi_{af}$, are important experimental probes of the nature of $\azz$ and $\fz$, as they are sensitive to the couplings in the processes of $\azz\to\kkbar$ and $\fz\to\kkbar$, respectively. A direct measurement of the mixing intensities would provide crucial constraints in models of $\azz$ and $\fz$ internal structure. It is also worth noting that besides the $\azz$-$\fz$ mixing mechanism, the underlying electromagnetic (EM) processes of $\jpsi\to\phi\azz$ and $\chi_{c1}\to\piz\fz$ with normal widths of $\azz$ and $\fz$ may occur via a $\gam^*$ or $K^*K$ loop~\cite{JJWu1}, and the EM processes will interfere with the corresponding mixing signals.

 As suggested by Refs.~\cite{JJWu1, JJWu2}, the mixing intensities, $\xi_{fa}$ and $\xi_{af}$, were measured based on the data samples of $2.25\times 10^8$ $\jpsi$ events and $1.06\times 10^8$ $\psip$ events collected at BESIII via the decays of $\jpsi\to\phi\eta\piz$ and $\chi_{c1}\to\pip\pim\piz$~\cite{YDWang}. At present, BESIII has accumulated much larger data samples of $1.31\times10^{9}$ $\jpsi$ events~\cite{njpsi09,njpsi12} and $4.48\times10^{8}$ $\psip$ events~\cite{npsip09,npsip12}, which provides a good opportunity to firmly establish the existence of $\azz$-$\fz$ mixing and precisely measure the mixing intensities. In this Letter, we present a study of $\azz$-$\fz$ mixing with the decays of $\jpsi\to\phi\eta\piz$ ($\eta\to\gam\gam$ and $\eta\to\pip\pim\piz$, $\phi\to\kpkm$) and $\psip\to\gamma\chico\to\gamma\piz\pip\pim$. The signals of $\azz$-$\fz$ mixing are observed with a statistical significance of larger than 5$\sigma$ for the first time, and the corresponding branching fractions and mixing intensities are measured.

 Details on the features and capabilities of the Beijing Electron-Positron Collider (BEPCII) and the BESIII detector can be found in Refs.~\cite{bepc,bes3}. A \textsc{geant}{\footnotesize 4}-based~\cite{geant4} Monte Carlo (MC) software package is used to optimize the event selection criteria, estimate backgrounds, and determine the detection efficiency. The selection criteria of charged tracks, particle identification (PID), and photon candidates are the same as those in Ref.~\cite{YDWang}.

 For the decay $\jpsi\to\phi\eta\piz$ with $\eta\to\gam\gam$ ($\pip\pim\piz$), the candidate events are required to have two kaons (and two pions) with opposite charges and at least four photons. A four-constraint (4C) kinematic fit enforcing energy-momentum conservation is performed for the $\kpkm(\pppm)\gam\gam\gam\gam$ hypothesis. For the events with more than four photons, the combination with the smallest $\chi^2_{\text{4C}}$ is retained, and $\chi^{2}_{\text{4C}}<50~(60)$ is required. For the $\eta\to\gam\gam$ decay mode, the $\piz$ and $\eta$ candidates are reconstructed by minimizing $\chi^2_{\piz\eta}= \frac{(M_{\gam_1\gam_2}-m_{\piz})^2}{\sigma_{\piz}^2}+\frac{(M_{\gam_3\gam_4}-m_{\eta})^2}{\sigma_{\eta}^2}$, where $m_{\piz}$ and $m_{\eta}$ are the nominal masses of $\piz$ and $\eta$~\cite{PDG}, $M_{\gam_1\gam_2}$ and $M_{\gam_3\gam_4}$ are the invariant masses of the $\gam_1\gam_2$ and $\gam_3\gam_4$ combinations, and $\sigma_{\piz}$ and $\sigma_{\eta}$ are their corresponding resolutions, respectively.
 The $\piz$ and $\eta$ candidates are required to satisfy $|M_{\gam_1\gam_2}-m_{\piz}|<15$~$\mevcc$ and $|M_{\gam_3\gam_4}-m_{\eta}|<30$~$\mevcc$, respectively. To reject the backgrounds of $\jpsi\to\kpkm\piz\piz$ and $\jpsi\to\kpkm\eta\eta$, two analogous chi-square variables $\chi^{2}_{\piz\piz}$ and $\chi^{2}_{\eta\eta}$ are defined, and candidates are required to satisfy $\chi^{2}_{\piz\piz}>40$ and $\chi^{2}_{\eta\eta}>5$. In the decay $\eta\to\pppm\piz$, the two $\piz$ candidates surviving the $\piz$ mass window requirement and with the smallest $\chi^2_{\piz\piz}$ are kept for further analysis. The $\pip\pim\piz$ combination with an invariant mass $M_{\pip\pim\piz}$ closest to $m_{\eta}$ and in the mass window of $|M_{\pi^+\pi^-\pi^0}-m_\eta|<20$~$\mevcc$ is regarded as the $\eta$ candidate. Finally, the $\phi$ signal events are identified by requiring $|M_{\kpkm}-m_{\phi}|<10$~$\mevcc$ for both $\eta$ decay modes, where $M_{K^+K^-}$ denotes the $K^+K^-$ invariant mass and $m_{\phi}$ is the $\phi$ nominal mass.

\begin{figure}[htbp]
   \centering
   \vskip -0.2cm
   \hskip -0.1cm \mbox{
   \begin{overpic}[width=4.3cm,height=3.5cm,angle=0]{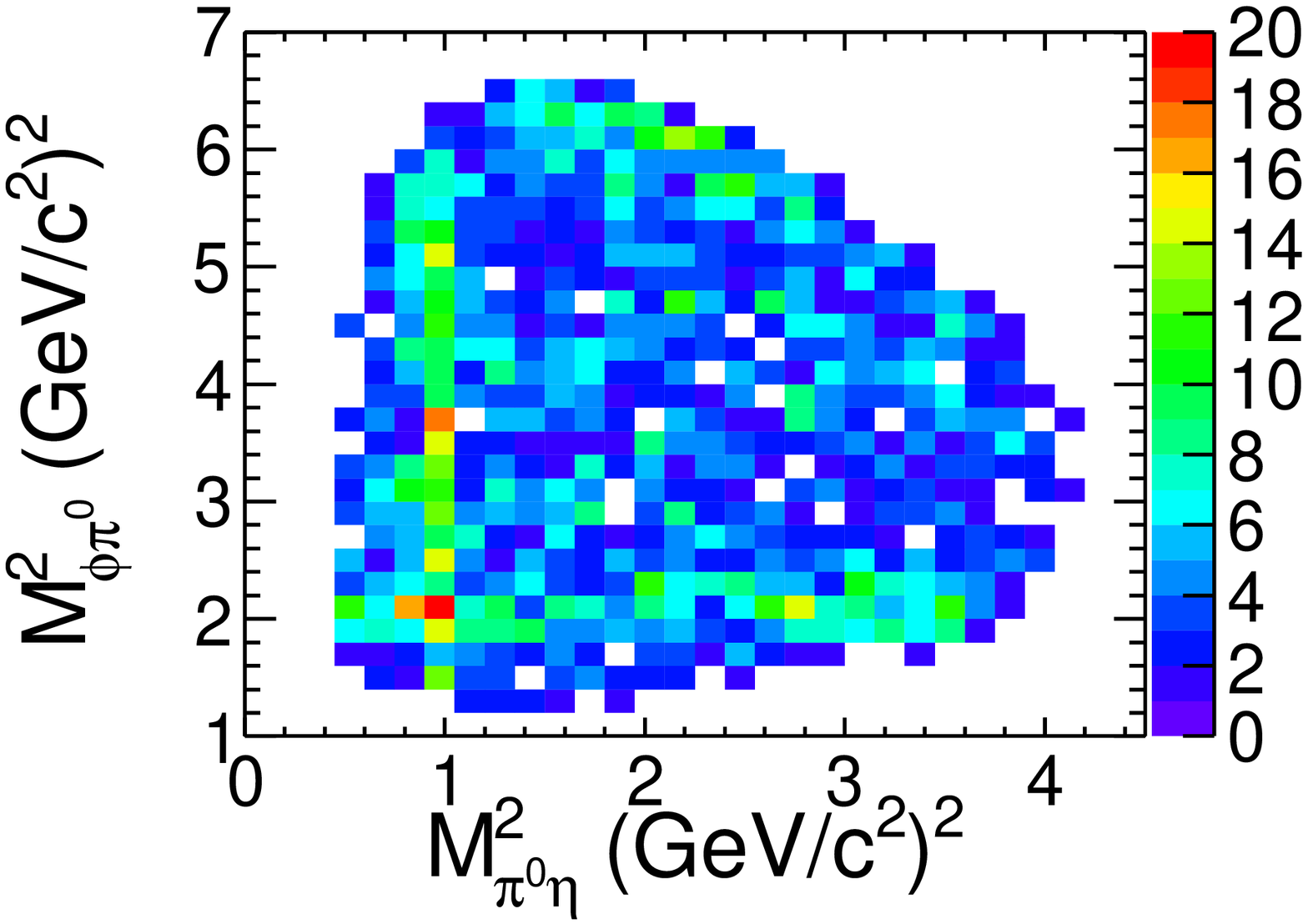}
   \put(22,67){\bf(a)}
   \end{overpic}
   \begin{overpic}[width=4.3cm,height=3.5cm,angle=0]{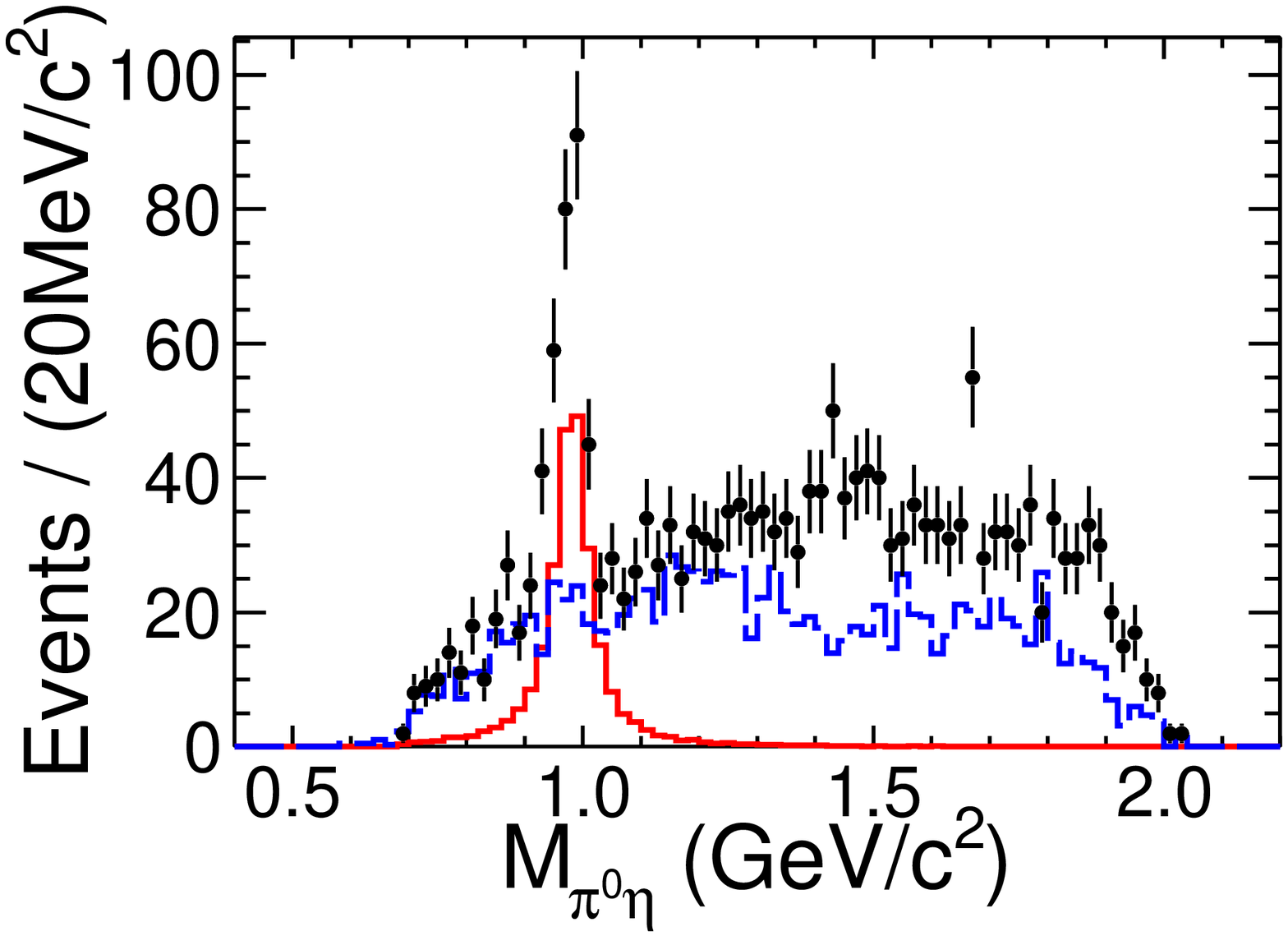}
   \put(22,67){\bf(b)}
   \end{overpic}}
   \vskip -0.0cm
   \hskip -0.1cm \mbox{
   \begin{overpic}[width=4.3cm,height=3.5cm,angle=0]{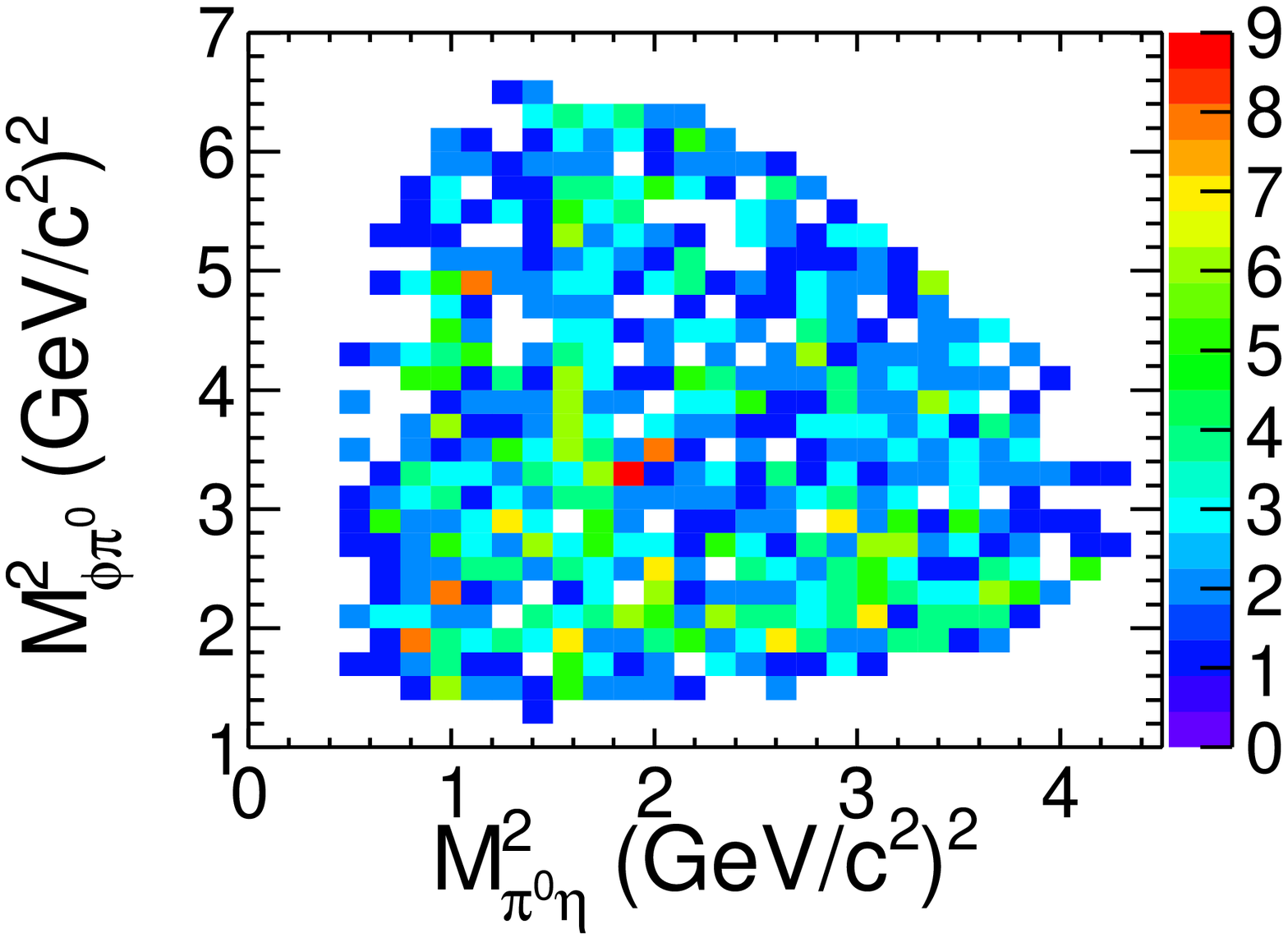}
   \put(22,67){\bf(c)}
   \end{overpic}
   \begin{overpic}[width=4.3cm,height=3.5cm,angle=0]{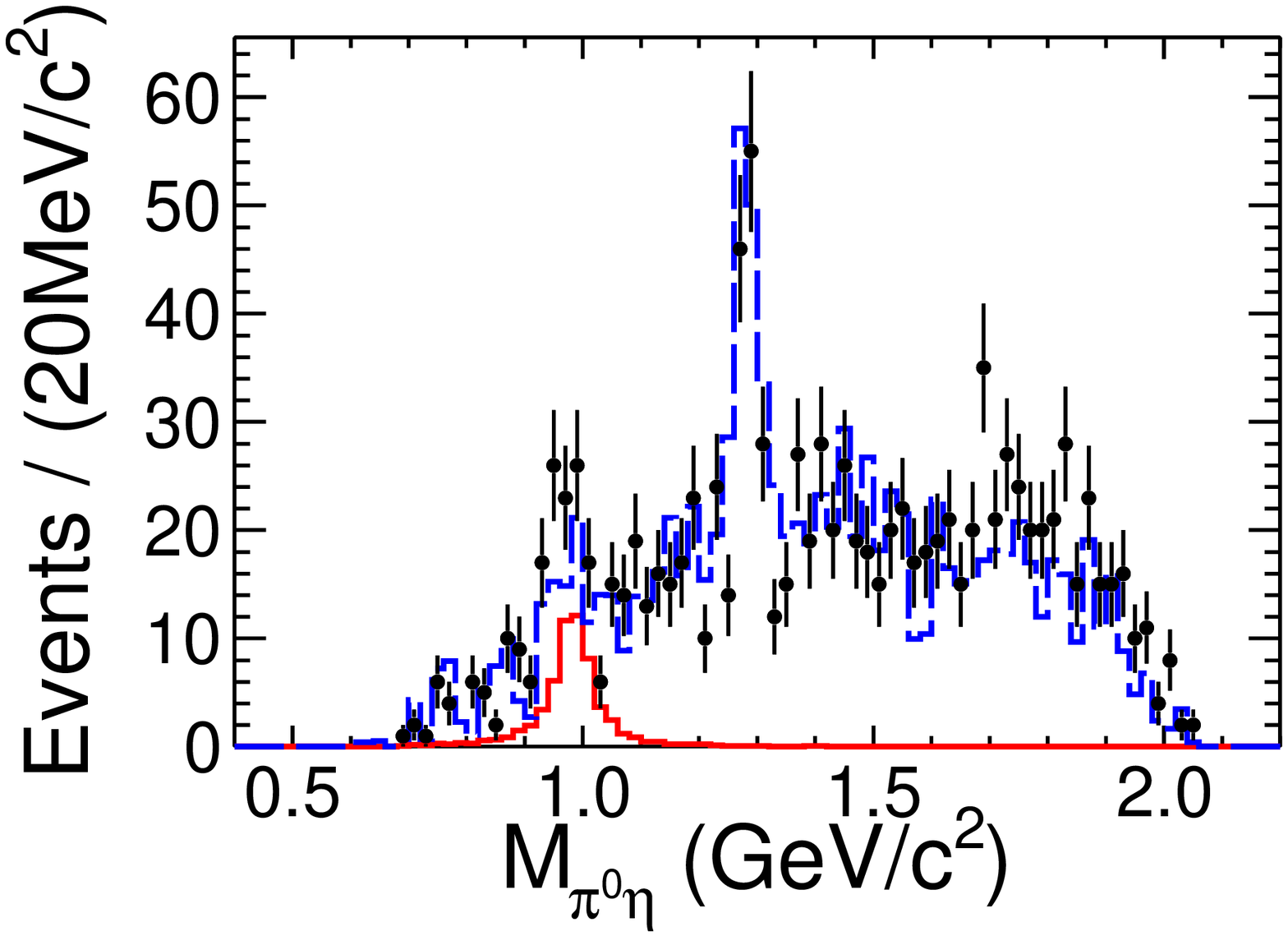}
   \put(22,67){\bf(d)}
   \end{overpic}}
   \vskip -0.1cm
   \hskip -0.1cm
   \caption{(color online) Dalitz plots for $\jpsi\to\phi\eta\piz$ (left) and mass projections on $M_{\piz\eta}$ (right); the upper-row plots are for the decay $\eta\to\gam\gam$, the down-row plots are for the decay $\eta\to\pip\pim\piz$. Dots with error bars represent the data, the (red) solid curves represent the MC simulated $\jpsi\to\phi\azz$ EM process, with the width of $\azz$ being set to its nominal value and amplitudes being normalized to the fit results, and the (blue) dashed curves represent the backgrounds estimated by the sideband events. }
   \label{a0_dalitz}
 \end{figure}

 After applying the above selection criteria, the Dalitz plots and the projections of $M_{\piz\eta}$ of the accepted candidate events of $\jpsi\to\phi\eta\piz$ are shown in Fig.~\ref{a0_dalitz} for the two $\eta$ decay modes. Prominent $\azz$ signals are observed. In the $\eta\to\pip\pim\piz$ decay, an obvious $f_{1}(1285)$ signal from the background $\jpsi\to\phi f_1(1285)$ ($f_{1}(1285)\to\pip\pim\piz\piz$) is observed in Fig.~\ref{a0_dalitz}(d). In addition, there are also horizontal bands around 2.0~$(\gevcc)^2$ in the Dalitz plots, which originate mainly from $\jpsi\to\eta K^{*\pm}K^{\mp}$ events within the $\phi$ mass window requirement. A detailed study indicates that the background events are dominantly from the non-$\phi$ and non-$\eta$ processes, while the non-$\piz$ backgrounds are negligible. Therefore, the background events in the $\phi$ sideband region ($1.05<M_{\kpkm}<1.07~\gevcc$) and $\eta$ sideband regions ($0.06<|M_{\gam\gam(\pppm\piz)}-m_{\eta}|<0.12~\gevcc$) are used to estimate the backgrounds. The expected backgrounds are shown as the dashed lines in Figs.~\ref{a0_dalitz}(b) and (d), and the $f_{1}(1285)$ peak is well described by the sideband events. A small $\eta^{\prime}$ signal is observed underneath the $\azz$ peak, which originates mainly from the decay $\jpsi\to\phi\eta^{\prime}$. To improve the mass resolution of the $\azz$ signal, a 6C kinematic fit with additional mass constraints on the $\eta$ and $\piz$ candidates is performed to the remaining events for the two $\eta$ decay modes, respectively, and the resulting $M_{\eta\piz}$ distributions are used to determine the signal yields.

 To determine the $\ftoa$ mixing signal in $\jpsi\to\phi\eta\piz$, an unbinned maximum likelihood method is used to simultaneously fit the $\piz\eta$ mass spectra for the two $\eta$ decay modes in the range of [0.70, 1.25]~GeV/$c^2$. In the fit, the $\ftoa$ mixing signal, the EM $\azz$ signal as well as their interference are considered as
 \begin{equation}
   |A_{\text{mix}}(m)\cdot e^{i\varphi} \cdot \alpha+ A_{a}(m)|^2 \times(p\cdot q),
   \label{a0_pdf}
 \end{equation}
 where $A_{\text{mix}}(m)=\frac{D_{fa}}{D_{a}\cdot D_{f}}$ is the amplitude of the mixing  signal~\cite{JJWu1, JJWu2}, $D_{a}$ and $D_{f}$ in the denominators are the $\azz$ and $\fz$ propagators, respectively, and $D_{fa}=\frac{g_{a_{0}K^{+}K^{-}} \cdot g_{f_{0}K^{+}K^{-}}} {16\pi}\times {i [\rho_{K^{+}K^{-}}(s)-\rho_{K^{0}\overline{K}^{0}}(s)]}$ is the mixing term. Here, $\rho_{K\bar{K}}(s)$ is the velocity of the $K$ meson in the rest frame of its mother particle, and $s$ is the square of center-of-mass energy of the mother particle. $A_{a}(m)=\frac{p^{L_{1}}q^{L_{2}}}{M^2_{a_0} - s - i\sqrt{s}[\Gamma^{a_0}_{\eta\piz}(s) + \Gamma^{a_0}_{K\overline{K}}(s)]}$ is a Flatt\'{e} formula for the EM $\azz$ signal.
 $\Gamma^{a_0}_{\eta\piz}(s)= \frac{g^{2}_{a_{0}\eta\piz}}{16\pi\sqrt{s}}\rho_{\eta\piz}(s)$ and $\Gamma^{a_0}_{K\overline{K}}(s)= \frac{g^{2}_{a_{0}\kpkm}}{16\pi\sqrt{s}}(\rho_{\kpkm}(s)+\rho_{K^{0}\overline{K^{0}}}(s))$ are the partial widths of $\azz\to\eta\piz$ and $\azz\to\kpkm$, respectively, where $g^{2}_{a_{0}\eta\piz}$ and $g^{2}_{a_{0}\kpkm}$ are the coupling constants and $p$ and $q$ are the momenta of $\azz$ and $\piz$ in the rest frames of $\jpsi$ and $\azz$, respectively. $L_{1}$ and $L_{2}$ are the corresponding orbital angular momenta and $\alpha$ and $\varphi$ represent the magnitude and relative phase angle, respectively, between the mixing signal and the EM process.

 \begin{figure}[htbp]
   \centering
   \vskip -0.2cm
   \hskip -0.0cm \mbox{
   \begin{overpic}[width=8.8cm,height=3.6cm,angle=0]{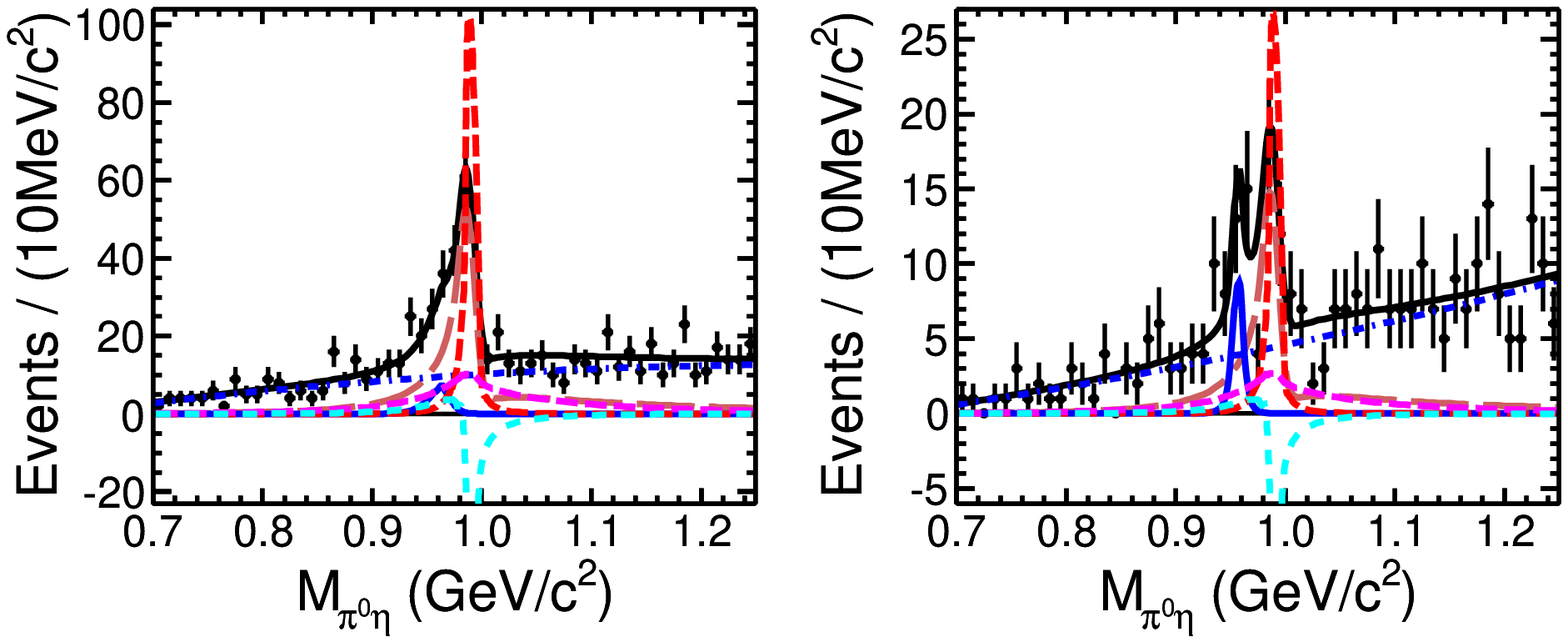}
   \put(12.5,33){\bf(a)}
   \put(61,33){\bf(b)}
   \end{overpic}}
   \vskip -0.0cm
   \hskip -0.0cm \mbox{
   \begin{overpic}[width=8.8cm,height=3.6cm,angle=0]{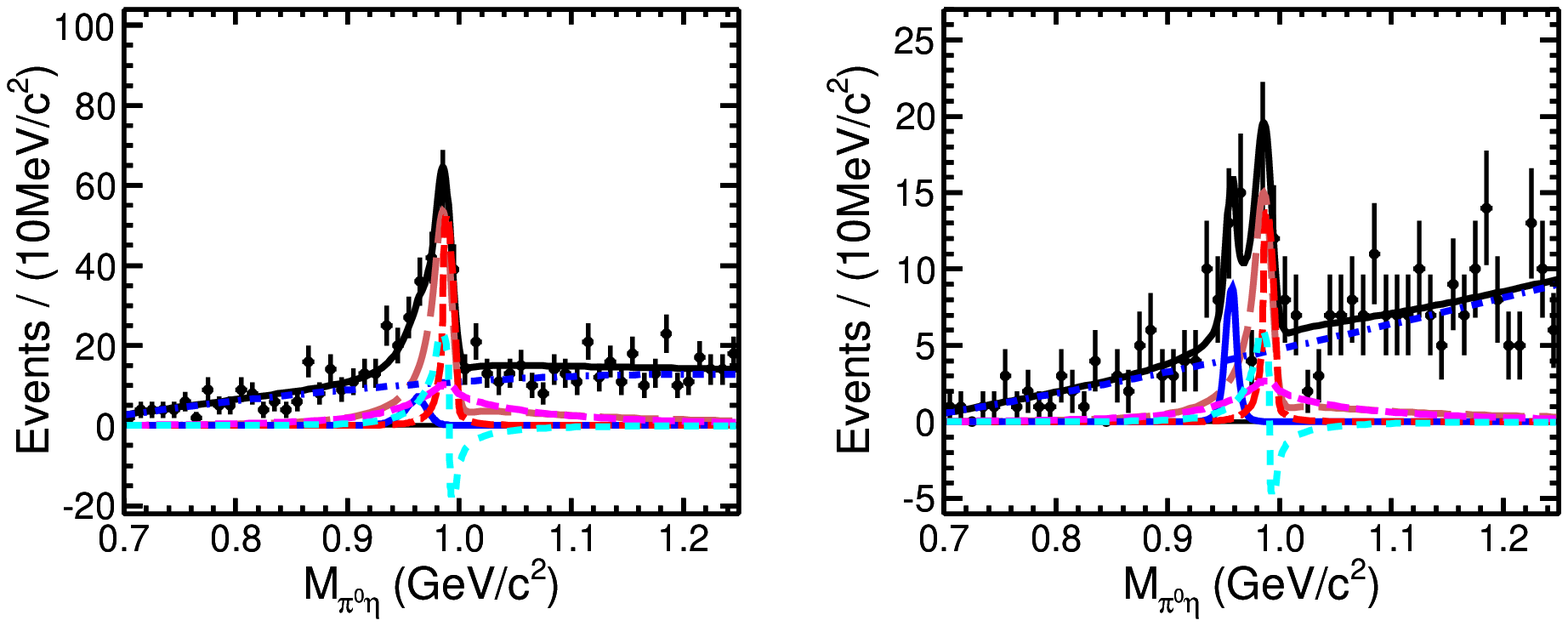}
   \put(12.5,33){\bf(c)}
   \put(61,33){\bf(d)}
   \end{overpic}}
   \vskip -0.1cm
   \hskip 0.0cm
   \caption{(color online) Fits to the $M_{\eta\piz}$ spectra of the $\jpsi\to\phi\eta\piz$ for destructive (upper) and constructive (lower) interference in the decay $\eta\to\gam\gam$ (left) and $\eta\to\pppm\piz$ (right), respectively. The dots with error bars represent the data, the (black) solid curves represent the total fit results, the (red) dashed curves represent the mixing signals, the (pink) dashed curves represent the $\jpsi\to\phi\azz$ EM processes, the (light-blue) dotted curves represent the interference terms, the (dark-red) long-dashed lines represent the sum of $\azz$ signals, the (blue) solid curves show the $\eta^{\prime}$ peaking backgrounds, and the (blue) dot-dashed curves represent the continuum backgrounds.}
   \label{a0_simufit}
 \end{figure}

 In the fit, the central masses and the coupling constants of $\azz$ and $\fz$ are fixed to the values obtained by the Crystal Barrel (CB) experiment~\cite{JJWu2,CB}. The mass resolution and the detection efficiency curve obtained from the MC simulation are taken into account. The two $\eta$ decay modes share identical parameters for the signal components in the fit. The background is represented by a second-order Chebyshev polynomial function with free parameters. The peaking backgrounds from the $\etap$ decays are included with shapes and magnitudes fixed to values estimated from the MC simulation. Two solutions (denoted as solution I and solution II for the destructive and constructive interferences, respectively) with different relative phase angles $\varphi$ but equal fit qualities are found. The statistical significances of the $\ftoa$ mixing signal and that of the $\jpsi\to\phi\azz$ EM process are $7.4\sigma$ and $4.6\sigma$, respectively, estimated by the changes of likelihood values between the fits with and without the mixing signal or EM process included. The resulting fit curves are shown in Fig.~\ref{a0_simufit}, and the signal yields are summarized in Table~\ref{yields}.

 For the decay $\psip\to\gam\chi_{c1}$, $\chi_{c1}\to\piz\pip\pim$, the candidate events are required to have two identified pions with opposite charge and at least three photons. A 4C kinematic fit is performed for the $\pip\pim\gam\gam\gam$ hypothesis. For events with more than three photons, the combination with the smallest $\chi^2_{\text{4C}}$ is retained, and $\chi^{2}_{\text{4C}}<20$ is required. To reject the background events with two or four photons in the final states, the two requirements $\chi^{2}_{\text{4C}}(\pip\pim\gam\gam\gam)<\chi^{2}_{\text{4C}}(\pip\pim\gam\gam\gam\gam)$ and  $\chi^{2}_{\text{4C}}(\pip\pim\gam\gam\gam)<\chi^{2}_{\text{4C}}(\pip\pim\gam\gam)$ are imposed. The $\piz$ candidate is reconstructed using the two-photon combination with invariant mass closest to $m_{\piz}$, and the same mass window is applied.

\begin{figure}[htbp]
   \centering
   \vskip -0.2cm
   \hskip -0.1cm \mbox{
   \begin{overpic}[width=4.3cm,height=3.5cm,angle=0]{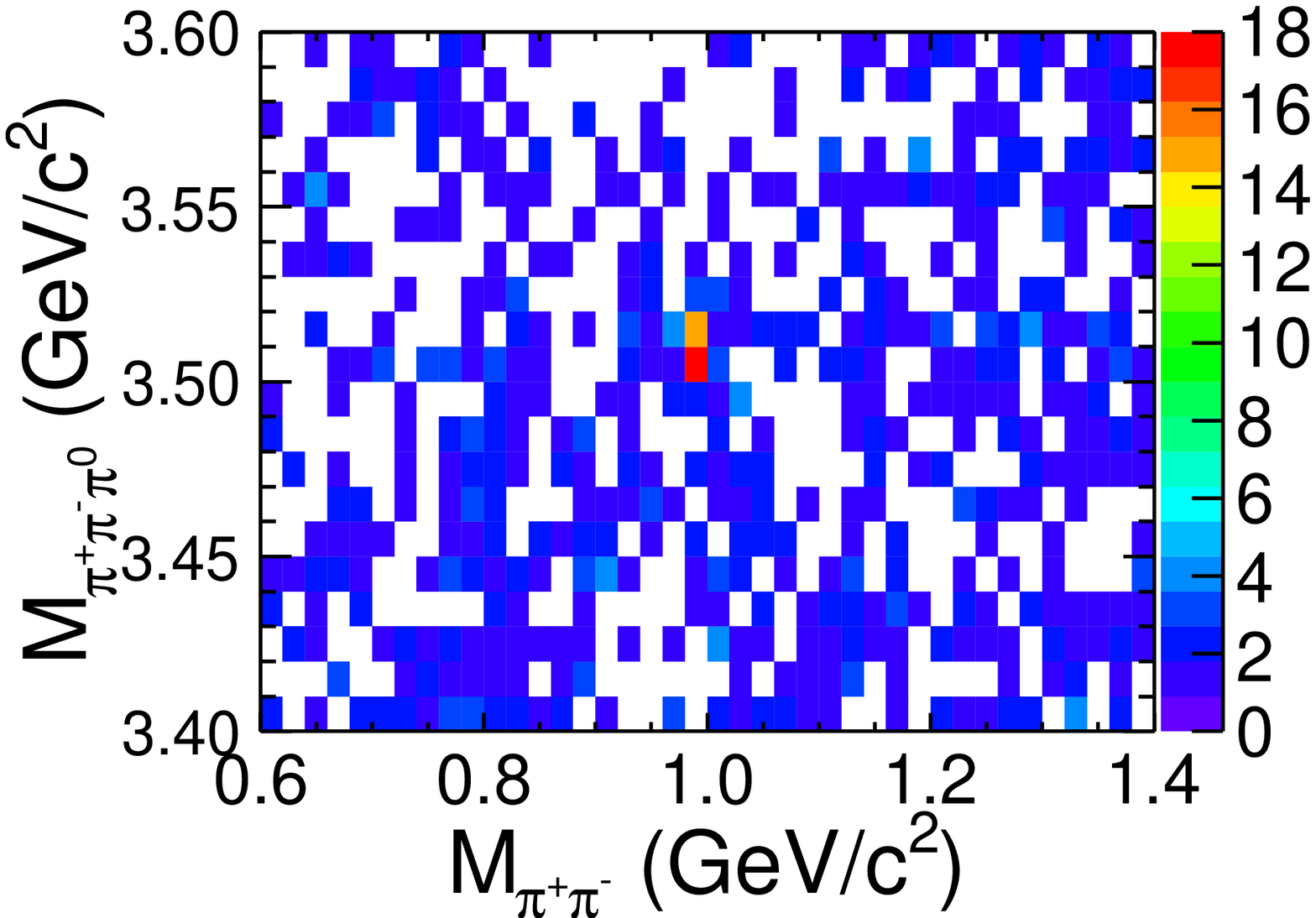}
   \put(22,67){\bf(a)}
   \end{overpic}
   \begin{overpic}[width=4.3cm,height=3.5cm,angle=0]{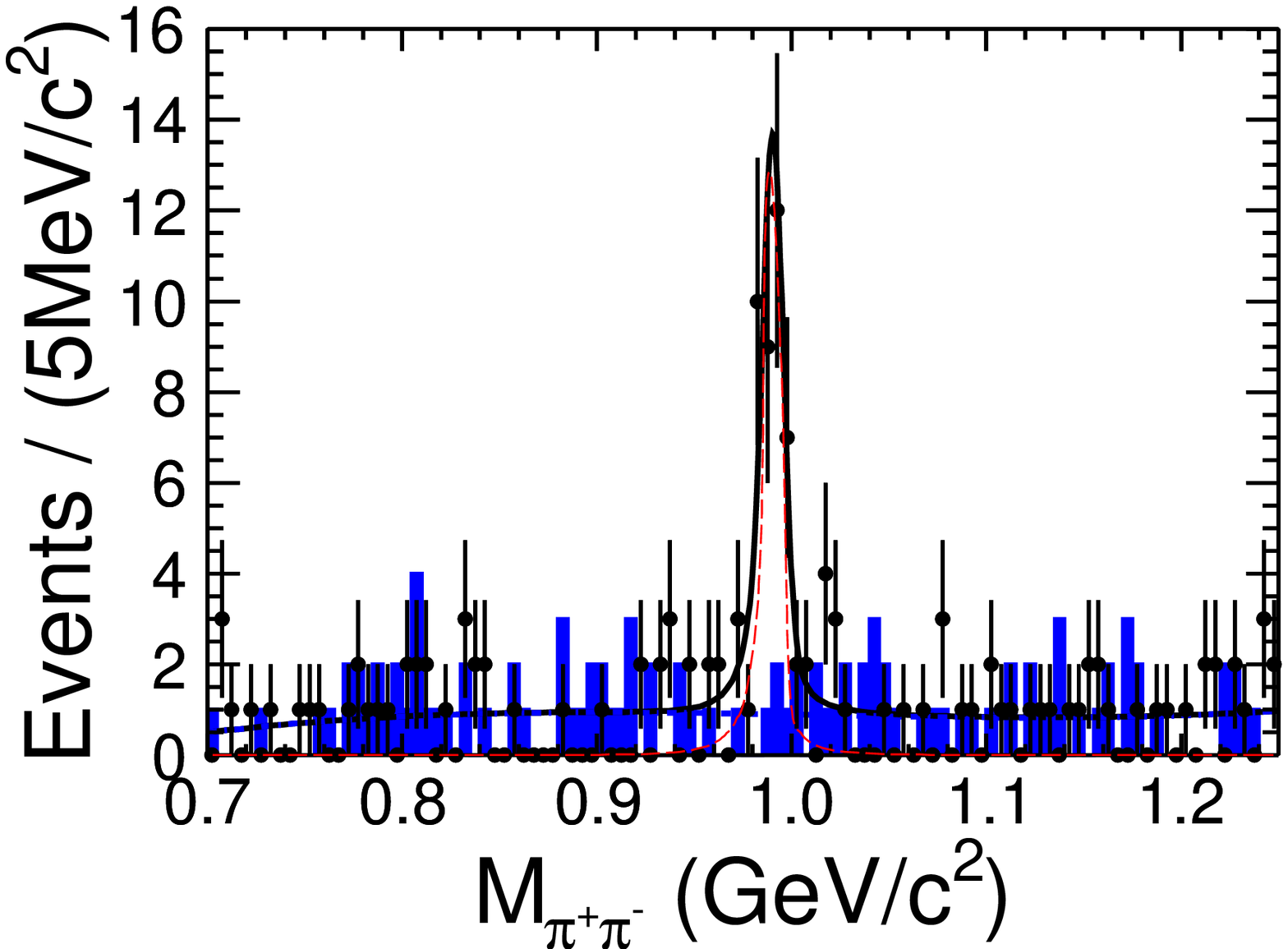}
   \put(22,67){\bf(b)}
   \end{overpic}}
   \vskip -0.1cm
   \hskip 0.0cm
   \caption{(color online) (a) Scatter plot of $M_{\pip\pim\piz}$ versus $M_{\pip\pim}$ for the $\chi_{c1}\to\pip\pim\piz$ decay and (b) fit to $M_{\pip\pim}$ spectrum for the $\chi_{c1}\to\pip\pim\piz$ in the $\chi_{c1}$ signal region. The dots with error bars are the data, the solid curve represents the fit result, the dashed curve represents the mixing signal, and the shaded histogram represents the normalized background from the $\chi_{c1}$ sideband. }
   \label{f0_simufit}
 \end{figure}

 After applying the above requirements, the scatter plot of $M_{\pip\pim\piz}$ versus $M_{\pip\pim}$ is shown in Fig.~\ref{f0_simufit}(a). A prominent cluster of $\chi_{c1}\to\piz\fz$ events is observed. The $M_{\pip\pim}$ projection with the $\chi_{c1}$ mass requirement of $|M_{\pip\pim\piz}-m_{\chi_{c1}}|<20$~MeV/c$^{2}$ is shown in Fig.~\ref{f0_simufit}(b). The width of the $\fz$ signal appears significantly narrower than the world average value~\cite{PDG}. The events from the $\chi_{c1}$ sideband region (3.43$<M_{\pip\pim\piz}<$3.47~GeV/$c^2$) and the inclusive MC sample are used to estimate the background shape, shown as the shaded histogram in Fig.~\ref{f0_simufit}(b), which is found to be flat in the $M_{\pi^+\pi^-}$ distributions.

 To determine the yield of the $\atof$ mixing signal in the $\chi_{c1}\to\pip\pim\piz$ decay, an unbinned maximum likelihood fit is performed to the $M_{\pip\pim}$ spectrum in [0.70, 1.25]~$\gevcc$. In the fit, the $\atof$ mixing signal and $\chi_{c1}\to\piz\fz\to\pip\pim\piz$ EM process are described in the same fashion as in Eq.(\ref{a0_pdf}), and the background shape is described by a second-order Chebyshev polynomial function. The fit result is illustrated in Fig.~\ref{f0_simufit}(b), and the signal yields are summarized in Table~\ref{yields}. The statistical significances of the mixing signal and the EM process are estimated to be 5.5$\sigma$ and 0.2$\sigma$, respectively. The interference effect between the mixing signal and EM process is weak enough to be neglected. The direct contribution from the EM process comes out to be negligible, and it is also ignored in the nominal fit. With the extracted signal yields, the branching fractions of the mixing processes $\jpsi\to\phi\fz\to\phi\azz\to\phi\eta\piz$, $\psip\to\gamma\chi_{c1}\to\gamma\piz\azz\to\gam\piz\fz\to\gamma\pppm\piz$ and the EM process $\jpsi\to\phi\azz\to\phi\eta\piz$, as well as the mixing intensities $\xi_{fa}$ and $\xi_{af}$, are calculated as summarized in Table~\ref{br sum}, where the normalization branching fractions are taken from the PDG~\cite{PDG}.

 \begin{table}[htbp]
  {\caption{Summary of the signal yields ($N$), relative phase angles ($\varphi$), and the statistical significance ($S$) from the fits, where the uncertainties are statistical only. In the decay $\jpsi\to\phi\eta\piz$ the former numbers are for the $\eta\to\gam\gam$ decay mode and the latter are for the $\eta\to\pip\pim\piz$ decay mode.}
  \label{yields}}
  \begin{small}
  \begin{tabular}{lccccccccc}
  \hline \hline
  \mulR{2}{*}{Channel}  &  \mulC{2}{c}{$\jpsi\to\phi\eta\piz$}        & \mulR{2}{*}{$\chi_{c1}\to3\pi$} \\
                        &~~~~{Solution I}~~~~~&~~~~~{Solution II}~~~~ &                  \\ \hline
    $N$\,(mixing)       &$161\pm26~|~45\pm7$  & $~67\pm21~|~19\pm6$   & $42\pm7$         \\
    $N$\,(EM)           &$162\pm54~|~46\pm16$ & $130\pm51~|~37\pm14$  &  $-$             \\
    $\varphi$\,(degree) &   $23.6\pm11.3$     &  $-51.5\pm21.3$       &  $-$             \\ \hline
    %N(inter)           &$-75\pm15|-21\pm4$   &   $23\pm5|6\pm1$      &  $-$             \\
    %N(total)           &$248\pm51|70\pm14$   & $220\pm49|62\pm14$    &  $-$             \\ \hline
    $S$\,(mixing)       &       \mulC{2}{c}{7.4$\sigma$}              & 5.5$\sigma$       \\
    $S$\,(EM)           &       \mulC{2}{c}{4.6$\sigma$}              &  $-$              \\
  \hline \hline
  \end{tabular}
  \end{small}
 \end{table}

 \begin{table*}[htbp]
  {\caption{The branching fractions ($\B$) and the intensities ($\xi$) of the $\azz$-$\fz$ mixing. The first uncertainties are statistical, the second ones are systematic, and the third ones are obtained using different parameters for $\azz$ and $\fz$ as described in the text.}
  \label{br sum}}
  \begin{small}
  \begin{tabular}{lccccccccc}
  \hline \hline
  \mulR{2}{*}{Channel}      &                   \mulC{2}{c}{$\fz\to\azz$}               &      \mulR{2}{*}{$\azz\to\fz$}    \\
                            &~~~~~~~~{Solution I}~~~~~~~~~&~~~~~~~~{Solution II}~~~~~~~ &                                   \\ \hline
  $\B$(mixing)~$(10^{-6})$   &~~~$3.18\pm0.51\pm0.38\pm0.28$~~~ & ~~~$1.31\pm0.41\pm0.39\pm0.43$~~~ &~~~$0.35\pm0.06\pm0.03\pm0.06$~~~      \\
  $\B$(EM)~$(10^{-6})$       & $3.25\pm1.08\pm1.08\pm1.12$ & $2.62\pm1.02\pm1.13\pm0.48$ &                ---                \\
  $\B$(total)~$(10^{-6})$   & $4.93\pm1.01\pm0.96\pm1.09$ & $4.37\pm0.97\pm0.94\pm0.06$ &                ---                \\
  $\xi$ (\%)                & $0.99\pm0.16\pm0.30\pm0.09$ & $0.41\pm0.13\pm0.17\pm0.13$ &  $0.40\pm0.07\pm0.14\pm0.07$      \\
  \hline \hline
  \end{tabular}
  \end{small}
 \end{table*}

 The systematic uncertainty for the branching fraction measurement mainly comes from uncertainties in the event selection efficiencies, the fit procedure, the branching fractions of intermediate state decays and the total numbers of $\jpsi$ and $\psip$ events. The uncertainties associated with the charged tracking and PID are both $1.0\%$ per track~\cite{track}, and $1.0\%$ for photon detection~\cite{photo}. For kinematic fits, differences in the efficiencies between data and MC are determined to be $1.5\%$ and $2.5\%$ by selecting clean control samples of $\jpsi\to\omega\eta\to\pip\pim\piz\eta$ and $\psip\to\pip\pim\jpsi\to\pip\pim\gamma\eta$, respectively. The uncertainties for $\phi$, $\eta$, $\piz$ and $\chi_{c1}$ mass window requirements are estimated as $1.8\%$, $0.1\%$, $1.0\%$ and $3.0\%$, respectively, while the contributions from the requirements on $\chi^{2}_{\piz\piz}$ and $\chi^{2}_{\eta\eta}$ are negligible. The uncertainty on the $\eta^{\prime}$ peaking background is estimated by varying $\eta^{\prime}$ yields by $1\sigma$ in the fit. The uncertainties on the continuum background shape are estimated as $3.4\%$ and $2.4\%$ for the two mixing processes by changing the order of the Chebyshev polynomial. The uncertainties on the branching fractions of the intermediate state decays are taken from PDG~\cite{PDG}. The uncertainties on the total numbers of $\jpsi$ and $\psip$ events are $0.8\%$~\cite{njpsi09,njpsi12} and $0.6\%$~\cite{npsip09,npsip12}, respectively. The total systematic uncertainties are the individual uncertainties added in quadrature (the correlation between the two $\eta$ decay modes in $\jpsi\to\phi\eta\piz$ is considered), as listed as the second item in Table~\ref{br sum}.

 Various experiments, \emph{e.g.}, BNL E852~\cite{BNL}, KLOE~\cite{KLOE1,KLOE2} and SND~\cite{SND1,SND2}, have reported different central masses and coupling constants for $\azz$ and $\fz$ resonances. To evaluate the likely impact from the input parameters of $\azz$ and $\fz$, a series of fits are carried out with the input masses and coupling constants from the different experiments. As the fit results turn out to be sensitive to the various input parameters, the largest deviations from the nominal results are treated as isolated uncertainties and are summarized as the third term in Table~\ref{br sum}.

\begin{figure}[htbp]
   \centering
   \vskip -0.2cm
   \hskip -0.1cm \mbox{
   \begin{overpic}[width=7.0cm,height=5.0cm,angle=0]{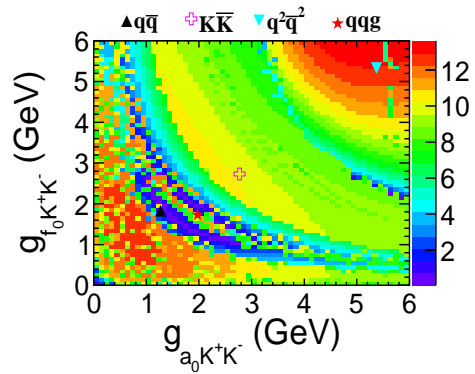}
   \end{overpic}}
   \vskip -0.1cm
   \hskip 0.0cm
   \caption{(color online) The statistical significance of the signal scanned in the two-dimensional space of $g_{a_{0}K^{+}K^{-}}$ and $g_{f_{0}K^{+}K^{-}}$. The regions with higher statistical significance indicate larger probability for the emergence of the two coupling constants. The markers indicate predictions from various illustrative theoretical models.}
   \label{signif}
 \end{figure}

 We obtain constraints on $g_{a_{0}K^{+}K^{-}}$ and $g_{f_{0}K^{+}K^{-}}$ by scanning the two coupling constants in the region of [0.0, 6.0]GeV, which covers all the results from theories and experiments, and calculate the statistical significance of the mixing signal by simultaneously fitting the $M_{\eta\piz}$ and $M_{\pip\pim}$ spectra in the data. Other input parameters of $\azz$ and $\fz$ are fixed to the CB-experiment values in the fit. The statistical significance of the signal versus the values of $g_{a_{0}K^{+}K^{-}}$ and $g_{f_{0}K^{+}K^{-}}$ is shown in Fig.~\ref{signif}. The regions with higher statistical significance indicate larger probability for the emergence of the two coupling constants. The predicted coupling constants from various models~\cite{JJWu2} are displayed as well (color markers), but the theoretical uncertainties on the models are not considered here.

 In summary, the $\azz$-$\fz$ mixing is studied with the isospin violating processes $\jpsi\to\phi\azz\to\phi\eta\piz$ and $\chi_{c1}\to\piz\fz\to\piz\pip\pim$ using the samples of $1.31\times10^{9}$ $\jpsi$ events and $4.48\times10^{8}$ $\psip$ events accumulated at the BESIII detector. Based on the input parameters of $\azz$ and $\fz$ in Refs.~\cite{JJWu2,CB}, the signals of $\ftoa$ and $\atof$ are observed for the first time with statistical significances of 7.4$\sigma$ and 5.5$\sigma$, respectively. The corresponding branching fractions of the mixing signal and the mixing intensities as well as the EM process of $\jpsi\to\phi\azz\to\phi\eta\piz$ are also measured. Finally, the significance of the mixing signal is measured versus the values of the two coupling constants, $g_{a_{0}K^{+}K^{-}}$ and $g_{f_{0}K^{+}K^{-}}$, and compared with theoretical predictions. In addition, the central values of the mixing intensities, $\xi_{fa}$ and $\xi_{af}$, can be used to estimate the coupling constants of the $f_0(980)$ and $a^0_0(980)$ resonances~\cite{xixi}. The results of this measurement help to deepen our understanding of the nature of the $\azz$ and $\fz$ mesons.

 We would like to thank J.~J.~Wu for very helpful discussions. The BESIII collaboration thanks the staff of BEPCII, the IHEP computing center and the supercomputing center of USTC for their strong support. This work is supported in part by National Key Basic Research Program of China under Contract No. 2015CB856700; National Natural Science Foundation of China (NSFC) under Contracts Nos. 11125525, 11235011, 11322544, 11335008, 11425524, 11375170, 11275189, 11475169, 11475164, 11175189, 11675183, 11675184, 11705006, 11735014; the Chinese Academy of Sciences (CAS) Large-Scale Scientific Facility Program; the CAS Center for Excellence in Particle Physics (CCEPP); the Collaborative Innovation Center for Particles and Interactions (CICPI); Joint Large-Scale Scientific Facility Funds of the NSFC and CAS under Contracts Nos. 11179007, U1532102, U1232201, U1332201; CAS under Contracts Nos. KJCX2-YW-N29, KJCX2-YW-N45; 100 Talents Program of CAS; National 1000 Talents Program of China; INPAC and Shanghai Key Laboratory for Particle Physics and Cosmology; German Research Foundation DFG under Contract No. Collaborative Research Center CRC-1044; Istituto Nazionale di Fisica Nucleare, Italy; Koninklijke Nederlandse Akademie van Wetenschappen (KNAW) under Contract No. 530-4CDP03; Ministry of Development of Turkey under Contract No. DPT2006K-120470; Russian Foundation for Basic Research under Contract No. 14-07-91152; The Swedish Resarch Council; U. S. Department of Energy under Contracts Nos. DE-FG02-05ER41374, DE-SC-0010118, DE-SC-0010504, DE-SC-0012069; U.S. National Science Foundation; University of Groningen (RuG) and the Helmholtzzentrum fuer Schwerionenforschung GmbH (GSI), Darmstadt; WCU Program of National Research Foundation of Korea under Contract No. R32-2008-000-10155-0.

\end{document}